\documentclass[intlimits,twoside,a4paper]{article}

\usepackage[utf8]{inputenc}

\usepackage{mathrsfs}
\usepackage{bbold}
\usepackage{float}
\usepackage{caption}
\usepackage{amsmath} 
\usepackage{epsfig, amsfonts} 
\usepackage{hhline}
\usepackage[none]{hyphenat}
\usepackage[textsize=tiny]{todonotes}
\usepackage[greek,english]{babel}
\usepackage{teubner}

\usepackage[eqsecnum]{cmpj3}

\usepackage{bm}

\newcommand{\be}{\begin{equation}}
\newcommand{\ee}{\end{equation}}
\newcommand{\bea}{\begin{eqnarray}}
\newcommand{\eea}{\end{eqnarray}}
\newcommand{\bem}{\begin{matrix}}
\newcommand{\eem}{\end{matrix}}
\newcommand{\nnb}{\nonumber}

\def\koppa{\hbox{\foreignlanguage{greek}{\coppa}}}
\def\smallkoppa{{\hbox{\foreignlanguage{greek}{\footnotesize\coppa}}}}

\issue{2024}{27}{3}{33602}
\doinumber{10.5488/CMP.27.33602}

\title{On a previously unpublished work with Ralph Kenna}
\author[R. Kenna, B. Berche]{ 
	\framebox{R. Kenna} \orcid{0000-0001-9990-4277}\refaddr{label1,label3},
 	B. Berche \orcid{0000-0002-4254-807X}\refaddr{label2,label3} }
\addresses{
\addr{label1} Centre for Fluids and Complex Systems, Coventry University, England
\addr{label2} Laboratoire de Physique et Chimie Th\'eoriques,  CNRS - Universit\'e de Lorraine, UMR 7019,
    	 Nancy, France 
\addr{label3} $\mathbb{L}^4$ collaboration, Leipzig-Lorraine-Lviv-Coventry, Europe
}

\date{Received December 01, 2023, in final form January 04, 2024}
\begin{document}
\maketitle
\sloppy

\begin{abstract}
This is part of an unpublished work in collaboration with Ralph Kenna. It was probably
not mature enough at the time it was submitted more than ten years ago and it was rejected by the editors, but some of the ideas 
had later been published partially in subsequent works. I believe that this ``draft'' reveals a lot about Ralph's enthusiasm and audacity and deserves to be published now, maybe as a part of his legacy.
  \keywords 
			universality, finite-size scaling, upper critical dimension
 \end{abstract}

\section*{My friend Ralph Kenna}
I am writing these words a few days after Ralph Kenna has passed away. Ralph was much more than a collaborator, he was my friend. We have had a long, fruitful, friendly, enriching scientific collaboration. His enthusiasm led him to produce new, original, and often revolutionary ideas. This is one of the elements of Ralph's personality. 
This truly sealed our friendship, together with our conception of Physics and the views of the world that we shared, through strong social and political commitment. 
\\ \vskip-3mm

\noindent \begin{minipage}[c]{0.45\linewidth}
\parindent=6mm  Our collaboration has left many ideas unresolved, drafts unpublished, some probably too daring and based on speculations that were too poorly supported. Still, I would like to submit to the editors of this journal one of these texts which dates from 2012, with Jean-Charles Walter as a third co-author. The paper was rejected, 
but I think that it is worth publishing as part of Ralph's legacy, along with some of the critical comments made at the time by the referees, that led to its rejection. 
 \end{minipage}
 \begin{minipage}[c]{0.54\linewidth}
 	 \vspace{1pt}
   \centering
   \includegraphics[scale=0.25]{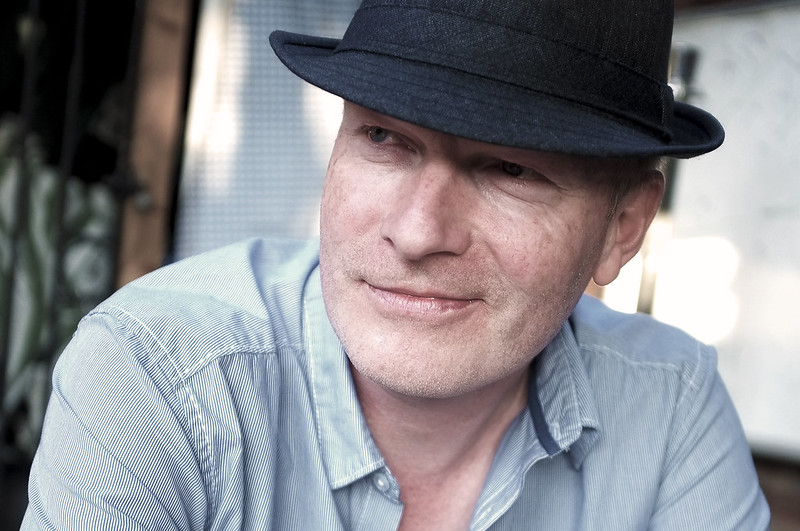}    
   \\ \small{Photo, courtesy of Thierry Platini.}
  \end{minipage}
 
 	 \vspace{0.5em}
This is not to circumvent the work of the referees. 
 The article included a hypothesis that was probably insufficiently supported, at least not enough to convince the referees, but it reflected much of the audacity of which Ralph was capable.
Ralph has always been unwavering in his support for Ukraine, in particular since the Russian invasion. The publication of one of his  bold scientific texts by a Ukrainian journal could be seen as a 
mutual, unfortunately, posthumous support.

%
%


\vskip0.5cm
\section*{The unpublished paper}\setcounter{section}{1}
I am writing these lines as if it were Ralph's work alone. This allows me to pay tribute to the fact that I considered him the driving force behind our collaboration. We were very complementary in our way of working, I believe. I sometimes held him back in speculations, and if I allowed myself a wink, even when he mixed Star Trek with our work!
The cover letter had been written by Ralph. His style can be recognized as his enthusiasm, using words like ``fundamental issues of statistical physics in high dimensions'', ``Our new theory incorporates or subsumes existing theories'' or ``shift in the paradigm'':
\begin{quotation}
The material in our paper connects with fundamental issues of statistical physics in high dimensions. We identify subtle, hidden flaws --- even at the level of mean-field theory --- which we believe have profound consequences. Because of the subtle nature of the issues we address, we offer here a very brief contextualization. (\dots) The statistical mechanics of condensed-matter, high-dimensional physics have been puzzling for a long time. It has been summed up by Kurt Binder et al. as ``a rather disappointing state of affairs'' --- ``the existing theories are not so good''. In addition, systems with free boundary conditions have been described by Peter Young et al. as particularly ``poorly understood''. Since they are experimentally accessible, the understanding of such systems impacts our understanding of finite-size materials with surfaces, comprising particles with long-range interactions.
Our new theory incorporates or subsumes the existing theories and is compatible with a vast amount of analytical and numerical evidence. However, our theory goes beyond this and introduces a powerful new principle that predicts and explains important features missed by current theories. For free boundaries, we show why 40 years of literature on the subject is based on an incorrect assumption.

We believe there is no current empirical or analytic evidence pointing against our new theory and it represents a shift in the paradigm of finite-size scaling and Landau mean-field theory in high dimensions. For these reasons, we would be grateful if you would consider the paper. 
\end{quotation}


\vskip0.5cm
\begin{abstract}
{\LARGE{\bf $Q$-information-entropic foundations of scaling}}
{\LARGE{\bf  in high dimensions and emergence of four }}\\
{\LARGE{\bf dimensionality}}\\ \\
{\normalsize
	\framebox{R. Kenna}$^1$,
 	B. Berche$^2$,
	J.-C. Walter$^3$
	}\\ \\
$^1$ Applied Mathematics Research Centre, Coventry University, England\\
$^2$ Statistical Physics Group, Institut Jean Lamour,  CNRS - Nancy Universit\'e - UPVM,     	 Nancy, France\\
$^3$ Instituut voor Theoretische Fysica, Katholieke Universiteit Leuven, Belgium  \\ \\
{March 22, 2012} \\

The emergence of a length scale that exceeds the linear extent of a finite-size, high-dimensional, pseudocritical system, and originates in the dangerous-irrelevant-variable mechanism of the renormalization group, is accompanied by a separation of associated dimensions. Here, we show that this phenomenon manifests a negative anomalous dimension associated with the correlation function alongside the effective, vanishing (Gaussian) one and allows universal extension of hyperscaling to high-dimensions as well as a second, fundamental fluctuation-response relation. We offer an information-entropic explanation for this phenomenon including for logarithmic corrections at the critical dimension. This mechanism offers a way to manifest short-range, high-dimensional Euclidean quantum-field systems as a four-dimensional without dimensional compact\-ification or sub-manifold restrictions.
\keywords{phase transition, critical exponents, upper critical dimension, hyperscaling relation, dangerous irrelevant variable}
 \end{abstract}
\date{March 22nd, 2012}



It is well known that standard finite-size scaling (FSS) is valid below the upper critical dimension $d = d_c$ when hyperscaling holds and where the correlation length is comparable to the linear extent of a system exhibiting a continuous phase transition \cite{1}. Above $d_c$, standard hyperscaling breaks down, and the bulk critical behaviour there is described by mean-field exponents \cite{2}. FSS was analyzed for $d \leqslant d_c$, Euclidean $\phi^4$ theory and the Ising model with periodic boundary conditions (PBC) \cite{3,4,5,6,7,8,9,10,10a,11,11a,11b,12,13,14,15,15a,15b,16,17} following large-$n$ analytical studies by Br\'ezin \cite{2}. The breakdown of standard FSS and hyperscaling is attributed to Fisher’s dangerous irrelevant variables \cite{18} in the renormalization-group (RG) framework \cite{3,4,5,6, 10,10a}. To repair FSS above $d_c$, Binder introduced another length scale that emerges from the RG treatment, dubbed the thermodynamic length \cite{3, 8, 13}. Below $d_c$, it coincides with the correlation length, while above $d_c$, it scales as a power of the system size. Extensive comparisons with numerical simulations have been performed and FSS above the upper critical dimension with PBCs is now considered to be well understood~\cite{11, 11a,11b, 12, 19, 20}. It is therefore perhaps surprising that, although the role of dangerous irrelevant variables in the breakdown of hyperscaling in high dimensions is well developed \cite{3,4,5,6,7,8,9,10,10a,11, 11a,11b,12,13,14,15,15a,15b,16,17}, FSS above the upper critical dimension was summarized by Binder et al. as ``a rather disappointing state of affairs --- although for the $\phi^4$ theory in $d = 5$ dimensions, all exponents are known, including those of the corrections to scaling, and in principle very complete analytical calculations are possible, the existing theories clearly are not so good'' \cite{14, 20}. In contrast to the PBC case, there have been relatively few studies of high-dimensional systems with free boundary conditions (FBC) \cite{7, 17}, which are complicated by additional scaling fields associated with boundaries in the RG picture \cite{19, 21}. The situation with FBCs was recently
described in reference~\cite{16} as ``poorly understood''.
Here, we present an alternative, corroborative theory and show that high-dimensional Ginzburg--Landau--Wilson physics is less ``trivial'' than hitherto realized. Although delivering some correct scaling and FSS behaviour above $d_c$, it is indeed lacking in several respects, most seriously for FBCs at the pseudocritical point. A comprehensive picture emerges by simply separating notions of underlying space and emergent space. The corresponding two notions of the correlation function, one of which has a stretched exponential form, are then associated with two separate anomalous dimensions and two associated fluctuation-response relations, only one of which is captured by mean-field theory. At the critical dimension, there are analogous pairs of logarithmic terms and scaling relations \cite{22, 23}. We demonstrate that existing analytic and numerical-based understandings of FBCs at pseudocriticality are unfounded and propose to postulate that FSS there is more similar to the PBC case than hitherto realized. Hyperscaling may then be extended beyond the upper critical dimension universally. After presenting numerical evidence supportive of our claims, we then propose an information-entropic foundation\footnote{[Footnotes are BB comments as of 2023.] This is probably the most audacious claim in the paper.} which lays behind, and greatly simplifies the dangerous-irrelevant-variables picture and delivers a new prediction for logarithmic corrections at 
$d = d_c$.

With the reduced temperature $t$ denoting the distance from the critical point, the standard, leading, critical scaling forms for the specific heat, spontaneous magnetization, susceptibility and correlation length are $c_\infty (t) \sim |t|^{-\alpha}$, $m_\infty (t) \sim |t|^{\beta}$, $\chi_\infty (t) \sim |t|^{-\gamma}$ and $\xi_\infty (t) \sim |t|^{-\nu}$, respectively. Here, the subscript indicates the linear extent of the system. At $t = 0$, the magnetization in field scales as $m_\infty(h) \sim |h|^{1/\delta}$. For sufficiently large distances $r$, and for $d < d_c$, the correlation function decays exponentially away from criticality $G(r) \sim r^{-p} \exp (-r/\xi)$ while at the critical point itself it reduces to a simple power law, $G(r) \sim r^{-(d-2+\eta)}$. Mean field theory predicts $p = (d - 1)/2$ outside the critical point and $\eta = 0$ at criticality. The six standard critical exponents $\alpha$, $\beta$, $\gamma$, $\delta$, $\nu$ and $\eta$ are related through the scaling relations. Of these, the hyperscaling relation
\be
\nu d=2-\alpha\quad \hbox{for}\quad d\leqslant d_c, 
\label{1.1}
\ee
is conspicuous in that it involves the dimensionality $d$ and fails above the upper critical dimension, which is $d_c = 4$ for the Ising model.
There, Landau’s mean field exponents are $\alpha = 0$, $\beta = 1/2$, $\gamma = 1$, $\delta = 3$, $\nu = 1/2$, $\eta = 0$.
Below $d_c$ dimensions, it is well known that the finite-size counterpart of the correlation length $\xi$ scales with the system extent $\xi_L\sim L$ and FSS is obtained by fixing the ratio $\xi_L/\xi_\infty(t)$ \cite{1}. Above $d = d_c$, one has instead that FSS is governed by\footnote{After a suggestion of M.E. Fisher, the exponent $q$ was later denoted in our subsequent papers as the archaic Greek letter  $\koppa$, see e.g. \cite{KBCMP13,EPL}. I adhere to this notation in the rest of this text and use $\koppa$ instead of $q$. The first 34 references are those of the original paper, those starting from~\cite{KBCMP13} are additional references available at the end of this article.}
\be 
\xi_{P_L}\sim L^q,\quad \hbox{where}\quad q=d/d_c,
\label{1.2}
\ee
which arises through dangerous irrelevant variables, at least for PBC’s \cite{2,3,4,5,6,7,8,9,10,10a,11,11a,11b,12,13,14}. The quantity $\xi_{P_L}$ has been dubbed characteristic length in reference~\cite{19} and has been related to the correlation length \cite{2, 16}, thermodynamic length \cite{3, 8} or coherence length \cite{19}. The emergence of the power-law relationship (\ref{1.2}) may be understood heuristically by demanding that the volume associated with $\xi_{P_L}$ (measured in units of $x_0$ in $d_c$ dimensions, say) should correspond to the actual volume of the system (measured in units of $z_0$ in $d$ dimensions),
\be 
\left( \frac{\xi_{P_L}}{x_0}\right)^{d_c}=\left(\frac{L}{z_0}\right)^d.
\label{1.3}
\ee
The replacement $\xi_\infty(t)\sim \xi_{P_L}$ yields $|t| \sim L^{-\smallkoppa/\nu}$, so that 
the susceptibility scales as
\be
\chi_L\sim L^{\frac{\gamma \smallkoppa}{\nu}}=L^{2\smallkoppa}.
\label{1.4}
\ee
 Equation~(\ref{1.3}) does not, however, capture the logarithms present in $d_c$ dimensions and these are addressed below. Evidence from a variety of studies supports equation~(\ref{1.4}) for PBC’s \cite{2,3,4,5, 7,8,9,10,10a,11,11a,11b,12,13,14,15,15a,15b,16,17}, but the prevailing picture is that $\koppa= 1$ for FBC’s \cite{7, 17, 24, 25}.

That $\koppa$ is indeed $d/d_c$ can be seen from the Lee-Yang zeros of the partition function, which offer another way to characterize phase transitions \cite{26,26a}. The standard scaling behaviour for the edge of their distribution is $h_{\rm edge}(t) \sim t^\Delta$ where $\Delta=\beta\delta$  is the gap exponent. The FSS for the $j$th zero is $h_j(L) \sim (j/L^d)^{\Delta \smallkoppa/\nu d}$ \cite{27}. Following reference~\cite{23,23a}, we write the finite-size susceptibility in
terms of the Lee-Yang zeros as 
$\chi_L\sim L^{-d}\sum_{j=1}^{L^d}h_j^{-2}(L)$
 which, at pseudo-criticality gives
\be
\chi_L\sim L^{\frac{2\Delta \smallkoppa}{\nu} -d}\sum_{j=1}^{L^d} j^{-\frac{2\Delta \smallkoppa}{\nu d}}(L).
\label{1.5}
\ee
Together with the standard, static scaling relations $2\beta+\gamma=2-\alpha$ and $\beta(\delta-1)=\gamma$, matching equation~(\ref{1.4}) to equation~(\ref{1.5}) gives the requirement that
\be 
\frac{\nu d}\koppa=2-\alpha.
\label{1.6}
\ee
This recovers the standard hyperscaling relation (\ref{1.1}) below $d_c$ dimensions provided $\koppa= 1$ there, as is well established. However, if $\koppa = 1$ persisted above $d_c = 4$ (in the Ising case), equation~(\ref{1.5}) would deliver a spurious leading logarithm in the $d = 6$ case, {\em irrespective of boundary conditions}. Since there is, in fact, no such logarithmic correction to the leading scaling behaviour \cite{2, 22, 23,23a, 28, 28a}, this indicates that $\koppa$ {\em cannot be 1 even for FBCs}. The incorporation of $\koppa$ into equation~(\ref{1.6}) essentially extends hyperscaling~(\ref{1.1}) beyond the upper critical dimension.

Having separated the notions of emergent length, volume, and dimensionality from those of the original system in equation~(\ref{1.3}), it is sensible to distinguish the associated spaces. We refer to the original $d$-dimensional system as $Q$-space, the $Q$-lattice or the $Q$-continuum and emergent, $d_c$-dimensional one as $P$-space, hence the notation $\xi_{P_L}$.

Fisher’s fluctuation-response relation is associated with the correlation function $G(r)$, which is also dimension dependent and needs reexamination to account for whether the distance is measured in the $d$-dimensional $Q$-space or the emergent $d_c$-dimensional $P$-space. When these are not distinguished (below the upper critical dimension), the standard derivation is to integrate $G$ over space giving
\be
\chi\sim\int_0^\xi G(r)r^{d-1}\rd r\sim\xi^{2-\eta},
\label{1.7}
\ee
leading to $\nu(2-\eta)=\gamma$.

Replacing $G(r)$ by unity in equation~(\ref{1.7}) gives the volume of
space to be $\int_0^{\xi_L} \rd r \,r^{d-1}=\xi_L^d$. This is correct below the 
upper critical dimension where $\xi_L\sim L$. Above $d = d_c$, however, bounding the integral by $\xi_{P_L}$ would erroneously give the volume of space to be $\xi_{P_L}^d\sim L^{d^2/d_c}$ . The problem is due to the failure to separate the notions of distance in $P$-space and $Q$-space. If the integral is bound by  $\xi_{P_L}$, one must integrate over the $d_c$ dimensions of $P$-space. Alternatively, if the integral is $d$-dimensional, it must be bound by  $\xi_{Q_L}\equiv  (\xi_{P_L})^{1/\smallkoppa}\sim L$. With $x$ referring to $P$-space distance, the former approach gives
\be
\chi_L\sim \int_0^{\xi_{P_L}}G_P(x) x^{d_c-1}\rd x\sim \xi_{P_L}^{2-\eta_P},
\label{1.8}
\ee
where $G_P(x) $ is the $P$-space correlation function
\be
G_P(x) \sim \frac{ \re^{-x/\xi_{P_L}} }{x^{(d_c-1)/2}}
\label{1.9}
\ee
away from criticality and
\be
G_P(x) \sim\frac{1}{x^{d_c-2+\eta_P}}
\label{1.10}
\ee
at it. This identifies the usual Fisher law
\be
\eta_P=2-\frac\gamma\nu
\label{1.11}
\ee
as a $P$-space relation. With $\gamma/\nu = 2$, we see that the mean field theory only captures the anomalous dimension of emergent $P$-space: $\eta_P = 0$. The $P$-distance $x$ is related to displacement $z$ in $Q$-space via $x \sim z^\smallkoppa$. In terms of this underlying scale, the counterparts of equations~(\ref{1.9}) and (\ref{1.10}) are
\be
G_Q(z) \sim\frac{\re^{-(z/\xi_{Q_L})^\smallkoppa}}{z^{(d-1)/2}}
\quad \hbox{and} \quad
G_Q(z) \sim\frac{1}{z^{d-2\smallkoppa}},
\label{1.12}
\ee
respectively.
Integrating this function over the $d$ dimensions of $Q$-space yields the correct FSS formula~(\ref{1.4}). We identify the anomalous dimension in fundamental $Q$-space as
\be
\eta_Q=2(1-\koppa)=2-\frac{\koppa\gamma}{\nu},
\label{1.13}
\ee
which is the fluctuation-response relation there.

Since $\alpha$ and $\nu$ in the extended version of hyperscaling (\ref{1.6}) are universal, $\koppa$ may be expected to be a new critical exponent. However, conventional wisdom has it that $\koppa = 1$ for FBC’s in particular, and that $\chi_L$ cannot diverge
more rapidly than $L^{\gamma/\nu}=L^2$ above $d_c$ \cite{4, 7, 10, 10a, 16, 24}.
In this sense, convention holds that FSS is not quite universal after all. Indeed, the failure to properly separate $\xi_{P_L}$ from $\xi_{Q_L}\sim L$ in equation~(\ref{1.7}), and in conventional FSS, would instead lead to the Gaussian form $\chi_L\sim L^2$.

\begin{figure}[ht]
\begin{center}
\includegraphics[width=8cm]{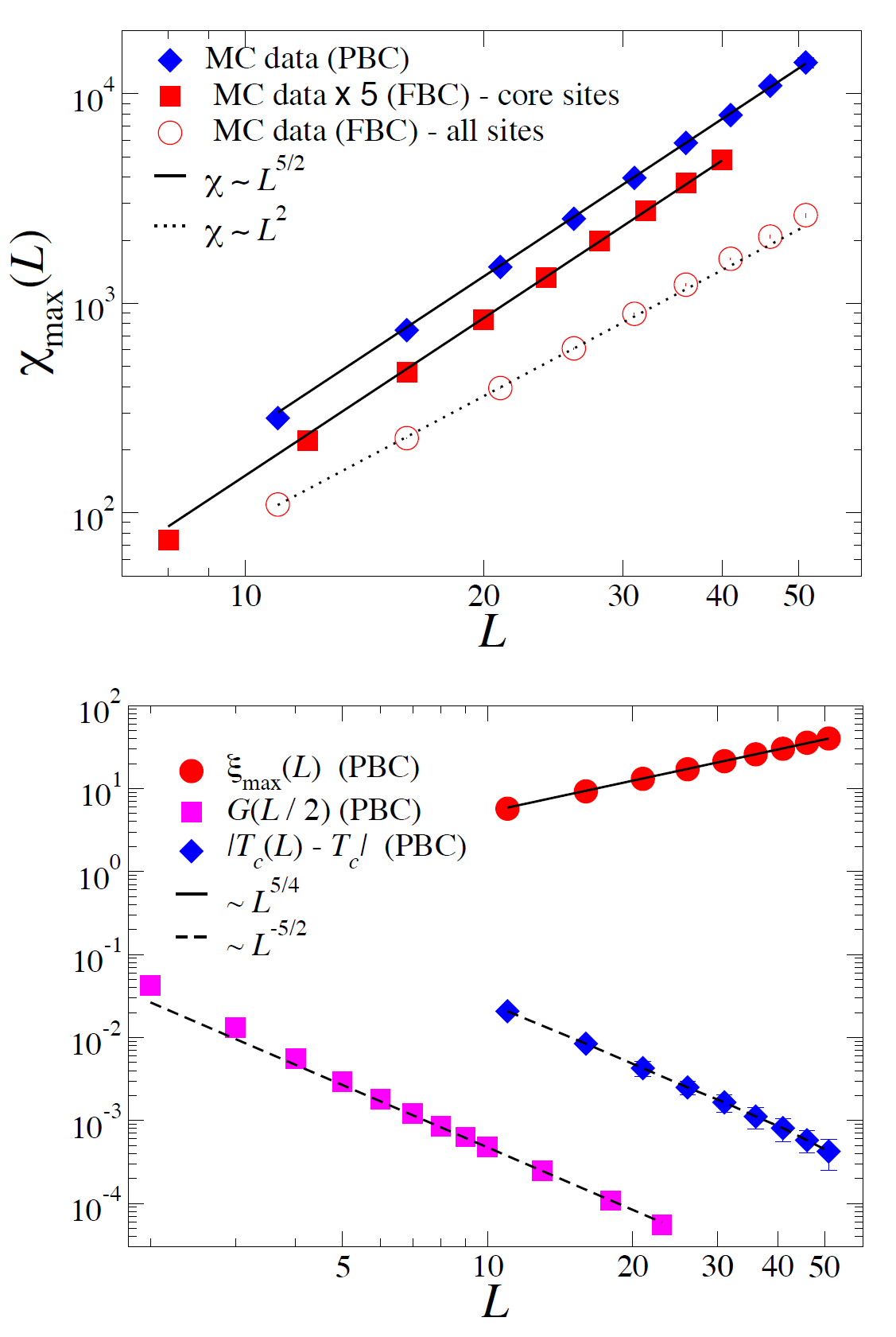}
\end{center}
\caption{(Colour online) 
(a) The susceptibility peak for 5D lattices with PBCs gives an effective exponent of 5/2 as predicted by $Q$-theory, rather than the Gaussian value 2. A similar plot for FBCs delivers the same result once boundary sites are removed. 
(b) The second-moment correlation-length peak, Lee-Yang edge, pseudocritical point and correlation function on 5D PBC lattices support the $Q$-theory predictions 
$\xi_L\sim L^\smallkoppa$, 
$h_1 \sim L^{-\smallkoppa\Delta/\nu}$, 
$h_c \sim L^{-3\smallkoppa}$, 
$t_L \sim L^{-\smallkoppa/\nu}$ and 
$G\sim L^{-(d-2\smallkoppa)}$
over conventional FSS (which is equivalent to $\koppa = 1$ in each case).
}
\label{figa}
\end{figure}

To test for universality, we simulated the Ising model in 5D using lattices with both PBCs and FBCs. In figure~\ref{figa}(a), the FSS of the susceptibility peak is plotted against $L$ in 5D and the form (\ref{1.4}) is verified, in agreement with references~\cite{2,3,4,5, 8,9,10,10a,11,11a,11b,12,13,14,15,15a,15b,16,17, 19, 20} in the PBC case.

For FBCs, the proportion of sites in the bulk of a
size-$L$ lattice is $(1 - 2/L)^d$, the remaining ones being in lower-dimensional manifolds. Thus, the $L = 4$ to $L = 20$, 5D lattices of the recent numerical work \cite{17} have only between 3\% and 59\% of sites in the bulk and do not represent five-dimensionality. The resulting conclusion is that $\chi_L\sim L^2$ is therefore not a 5D one. On the theoretical side, it is reported in reference~\cite{7} that equation~(\ref{1.4}) ``cannot hold for free boundary conditions because it lies above a strict upper bound'' (namely $L^{\gamma/\nu}=L^2$) established in reference~\cite{24} (see also reference~\cite{25}). However, that upper bound was determined at $T_c$, not at $T_L$, and using the fluctuation-response relation (\ref{1.11}) instead of (\ref{1.13}). Moreover, the Fourier analysis of reference~\cite{7}, which yielded the same conclusions, neglects the quartic part of the Ginzburg--Landau-Wilson $\phi^4$ action because of an expectation that the Gaussian result should ``apply as an exact leading order result in more than four dimensions''. It was shown in references~\cite{11,11a,11b} that for PBC, the anomalous FSS behaviour (\ref{1.4}) is obtained from precisely this term. These observations, together with the conclusion that $\koppa = 1$ delivers leading logarithms in $d = 6$ from equation~(\ref{1.5}), indicate that the FSS paradigm from numerical and analytical conclusions in over 40 years of literature on the susceptibility of FBC lattices above $d_c$ is unsupported, at least at pseudocriticality~\cite{7, 17, 24, 25}.

We simulated 5D FBC Ising lattices up to $L = 40$. To probe the $d$-dimensionality of the system, we remove the contributions of sites close to the surfaces resulting in lattice cores of size $L/2$. The plot demonstrates that this procedure changes the apparent effective critical exponent from 2 (coming from the 4D surface sites and erroneously hinting at Gaussian behaviour) to 5/2. (In fact in figure~\ref{figa}(a), the susceptibility peaks for FBC lattice cores are multiplied by 5 to bring them within the range of the plot, but this does not affect scaling.) This supports the universality of $\koppa$ and modified FSS at pseudocriticality.

To numerically extract the correlation length, we use the second moment of the correlation function~$\xi_L$. The FSS of the correlation-length peak and the pseudocritical point are given in figure~\ref{figa}(b) for PBC’s and verify the 5D scaling forms $\xi_L\sim\xi_{P_L}\sim L^{5/4}$  and $t_L \sim L^{-5/2}$ over the Gaussian forms $\xi_L\sim L$ and $t_L \sim L^{-2}$. The validity of modified FSS above $d_c = 4$ is further confirmed for the Lee-Yang edge, which scales as $h_1(L) \sim L^{-15/4}$ instead of $h_1(L) \sim L^{-3}$. The pseudocritical field scales as
$h_c (L) \sim L^{-15/4}$ instead of as $L^{-3}$. (The Lee-Yang zeros are multiplied by $10^4$ and the pseudocritical field by 200 to bring them within the range of the plot.) The form $ G \sim L^{-5/2}$ from equation~(\ref{1.12}) is also verified and the Gaussian form $G \sim L^{-2}$ is dispelled (see also reference~\cite{11,11a,11b}). This represents a fundamental change in our understanding of the behaviour of the correlation function in high dimensions --- the hitherto widely accepted mean field value of zero for the anomalous dimension is an effective one, holding when the distance is measured in emergent $P$-space only. In terms of the more fundamental distance scale of $Q$-space, the anomalous dimension is negative and given by equation~(\ref{1.13}).

To summarize so far, above the upper critical dimension, a second notion of distance $\xi_{P_L}\sim L^\smallkoppa$ emerges alongside 
$\xi_{Q_L}\sim L$ \cite{2, 3, 5, 8, 16, 19}. Each length scale has an associated dimensionality; $d$ for the fundamental $Q$-space and $d_c$ for emergent $P$-space. Correlation decay is governed by the stretched exponential (\ref{1.12}) in $Q$-space and the more usual form (\ref{1.9}) in $P$-space. Defining  $\nu_P=\koppa\nu_Q=\nu$  restores the new hyperscaling and fluctuation-response relations (\ref{1.6}) and (\ref{1.13}) to the standard forms $\nu D = 2-\alpha$ and $\nu(2-\eta)=\gamma$, provided $(D,\nu,\eta) = (d,\nu_Q,\eta_Q)$ in $Q$-space and $(D,\nu,\eta) = (d_c,\nu_P,\eta_P)$ in $P$-space. The thermodynamic limit is then characterized by 
$\xi_{Q_\infty}\sim |t|^{-\nu_Q}$
and $\xi_{P_\infty}\sim |t|^{-\nu_P}$  in fundamental and emergent space and only the latter is captured by mean-field theory. Thermodynamic functions $c$, $m$ and $\chi$ and associated exponents $\alpha$, $\beta$, $\gamma$, $\delta$ are the same in $Q$-and $P$-space since, rather than notions of length or dimensionality, they involve sums over the lattice or integrals over the continuum.

We now turn to logarithmic corrections at the upper critical dimension itself and propose a deeper reason behind the heuristic arguments for equation~(\ref{1.2}) given earlier. FSS at the upper critical dimension exhibits multiplicative logarithmic corrections of the form \cite{2, 16, 22, 23} 
\be
\xi_L\sim L(\ln L)^{\hat \smallkoppa},
\label{1.14}
\ee
which is not captured by the heuristic argument associated with equation~(\ref{1.3}). An empirical observation, which to our knowledge has gone unnoticed in the literature\footnote{An argument was later given in \cite{JJRL17}.}, is that the relationship
\be 
\hat \smallkoppa = \frac 1{d_c}
\label{1.15}
\ee
appears to hold for models at their upper critical dimensions: 
$d_c = 4$ for the Ising and $O(n)$ models for all values of $n$; 
$d_c = 6 $ for $m$-component spin glasses for all $m$, as well as for percolation and the Yang-Lee edge problem; and $d_c = 2\sigma$ for models with long-range interactions characterised by $\sigma$ \cite{23,23a}. Models below the upper critical dimension which exhibit logarithmic corrections, on the other hand, have $\hat \koppa = 0$ (e.g., the 4-state Potts and Ising models in $d = 2$, the random-bond or site Ising model in $d = 2$, and the $n$-color Ashkin-Teller model) \cite{23,23a}.

A transformation of the form $x = z^\smallkoppa$ from $Q$-space to
$P$-space is not bijective and is therefore associated with a
loss of information, which should be taken into account.
According to the Landauer Principle, any such logically
irreversible transfer of information must be accompanied
by an entropy increase \cite{29}. Recent experiments verify
that information is indeed physical and the conversion
of information to energy is possible \cite{30,30a}. In statistical
mechanics, information is measured through Shannon or
Hartley entropy, which is an extensive concept provided
the system under consideration has short-range interactions. The Hartley information content of $Q$-space is $H =
\ln W$ where $W = (L^d)!$ is the number of ways to place $L^d$
spins on the $Q$-lattice. Assuming the information loss is
proportional to the amount of information available, Landauer's theory predicts the energy gain in mapping from $Q$-
to $P$-space is  $\xi_{P_L}^{d_c} e_{P_L}-L^d e_{Q_L}=\ln (L^d !)=L^d\ln L^d$  by
Stirling’s approximation. Here, $e_{Q_L}$ and $e_{P_L}$ are the internal energies in $Q$-and $P$-space, respectively, and, dominated by the regular part, are constant to the leading order. Above the upper critical dimension, we have seen that we must account for long-range correlations since $\koppa > 1$ and $\eta_Q < 0$ in $Q$-space. We, therefore, promote the Hartley information entropy to Tsallis’s Q-logarithmic form, so that
\be
a\frac{\xi_{P_L}^{d_c}} {L^d}-1\approx b\ln_{\tilde q} L^d,
\label{1.16}
\ee
for constants $a$ and $b$. The $Q$-logarithm, defined as $\ln_{\tilde q}z  = (z^{\tilde q-1} - 1)/(\tilde q-1)$, becomes a usual logarithm in the limit $\tilde q\to 1$. The identification $\tilde q = 1/\koppa$ yields $\xi_{P_L} \sim   L^\smallkoppa (1 + cL^{ -(d-d_c) } )$, recovering equation~(\ref{1.3}) as the dominant behaviour when $d > d_c$ together with the same corrections as those from the RG treatment of the susceptibility in references~\cite{11,11a,11b}. If $d < d_c$, equation~(\ref{1.16}) delivers $\xi_{P_L}\sim L$ with corrections which are swamped by the Wegner irrelevant-field terms \cite{32,32a,32b}. In addition, the $\koppa = 1$ limit delivers equation~(\ref{1.15}), explaining the observations made earlier for various models. This also explains why $\hat \koppa = 0$ in models away from their upper critical dimensions \cite{23,23a}.

When $d = d_c$, the $P$- and $Q$-space correlation functions (\ref{1.10}) and (\ref{1.12}) become $G_P(x) \sim x^{-(d_c-2+\eta)} (\ln x)^{\eta_P}$ and $G_Q(x) \sim z^{-(d_c-2+\eta)} (\ln z)^{\eta_Q}$, respectively. With $\xi_\infty(t)\sim |t|^{-\nu}(- \ln |t|)^{\hat \nu}$
 and 
$ \chi_\infty(t)\sim |t|^{-\gamma} (-\ln |t|)^{\hat \gamma}$, logarithmic analogues to (\ref{1.11}) and (\ref{1.13}) are   $\hat\eta_P =\hat\gamma-\hat \nu (2-\eta)$ and
$\hat\eta_Q =\hat\gamma-(\hat \nu-\koppa) (2-\eta)$ 
 respectively (also see supplemental material).
In condensed matter physics, systems with long-range interactions have a lowered upper dimensionality \cite{12, 33}. Thus three, or even two-dimensional systems may reside above $d_c$ and be experimentally accessible. In the RG approach to such systems, the lattice spacing has physical meaning and the physical quantities are the bare ones. Similarly, in condensed matter, the $Q$-space introduced here is physical, while $P$-space is a mathematical construct without direct physical meaning. This is why the anomalous dimension is given by the new equation~(\ref{1.13}) rather than equation~(\ref{1.11}).

In high-energy, fundamental-particle systems, on the other hand, the lattice spacing is an artificial construct that regularizes quantum field theories. Continuous phase transitions, where the correlation length (inverse mass gap) diverges, are the portals through which such theories return to the continuum limit. Physics takes place on the length scale $\xi_L=\xi_{P_L}$ and the RG philosophy is that the emergent (renormalized) quantities are the physical, observable ones. Similarly, in $Q$-theory, $P$-space is physical spacetime which emerges via $Q$-information-entropic, dangerous irrelevancy \cite{4, 11,11a,11b, 12, 18} through Binder’s thermodynamic or correlation length \cite{3, 4, 8, 13} (see supplemental material). Thus, a $d_c = 4$-dimensional physical system, with length scale $\xi_{P_L}$, emerges from a Euclidean field theory with length scale $L$ and arbitrary dimensionality $d > 4$, while the anomalous dimension is the mean-field one given by equation~(\ref{1.11}).

Finally, in discussing spontaneous symmetry breaking, Coleman famously remarked that ``a little man living inside \dots a ferromagnet would have a hard time detecting the rotational invariance of the laws of nature'' \cite{34}. We now realize that he is confronted with a similar difficulty if he attempts to measure the dimensionality of his world. Since physics takes place on a scale associated with the emergent length, any attempt to measure the extent of a high-dimensional universe yields $\xi_{P_L}$ rather than the underlying length $L$ and the little person’s world will appear to have dimensionality $d_c = 4$ rather than $d$. Although we start with a highly correlated ($\nu_Q = \koppa\nu_P$ , $\koppa > 1$, $\eta_Q < 0$) $d > 4$ dimensional $Q$-space, the physical universe necessarily emerges as a Gaussian ($\nu_P = \nu = 1/2$, $\eta_P = 0$) four-dimensional one. The extension of hyperscaling to above four dimensions through the $Q$-information-entropic, dangerous-irrelevant mechanism proposed here delivers an explanation for the four-dimensionality of our physical universe without the need for dimensional compactification or submanifold restrictions which are features of other high-dimensional theories. $Q$-theory, therefore, provides a mechanism for hiding the extra dimensions that have long been suspected.

\vspace{1cm}
\noindent
{\bf{Acknowledgements:}} 
We thank Christophe Chatelain and Yurij Holovatch for discussions. This work is supported by the EU FP7 IRSES Projects 269139 and 295302.


\section*{The critiques from the referee}\setcounter{section}{2}
We received three referee reports. Two of them (say A and B) suggested the paper be submitted to a more specialized journal, stating it was not of sufficient general interest. B agreed that ``the issue of FSS for $d > 4$ certainly requires better understanding'' and that our ``ideas and claims might ... constitute progress''. C said we ``raise some interesting issues about scaling'' and expose ``an inconsistency which may not have been pointed out before''.
All this is rather positive (we all know that the general interest issue is referee-dependent).

\subsection*{The problem of ``Q-entropy''}

More serious is that
both Referees A and C were puzzled by our derivation of equation~(\ref{1.16})
and our usage of the
term ``Q-entropy'' and Referee B was concerned that our paper lacked a ``solid foundation in terms of a viable theory''. Also, Referee C asked ``Surely if I live in a seven-dimensional world, I
would be able to look out in all seven dimensions, and not think that I was living in only four dimensions?''.

As we see, 
the major criticism concerns the very subject of the article, and although
more details were available in the Supplemental Material that we had provided in support of our submission, this did not convince the Referees. We gave up and did not submit the same paper anywhere else, since we agreed eventually our hypothesis was not solid enough.

Nevertheless, I think it is useful now to provide an answer, using other insights from unpublished material.
We based the central hypothesis using a nonextensive entropic start. We were aware that nonextensive entropy, while enjoying the support of some famous names (Gell-Mann~\cite{GellMann}) is not universally supported (Nauenberg~\cite{Nauenberg}).
Nonextensive entropy is a generalization of the traditional Boltzmann-Gibbs entropy used in statistical mechanics. It has found applications in various fields, including complex systems, systems with long-range interactions which depart from extensivity, and non-equilibrium statistical mechanics (for example the volume from the Sante Fe Institute on the Sciences of Complexity edited  by Gell-Mann  and Tsallis~\cite{GellMann} is devoted to applications in complex systems).
On the other hand, nonextensive entropy
has faced several critiques. One of the main critiques revolves around its theoretical foundations and the departure from the standard principles of statistical mechanics.

Despite these caveats, since this single hypothesis is capable of explaining (i) the coincidence of the finite-size correlation length with system size below the upper critical dimensionality (ii) the power-law scaling $\xi=L^\smallkoppa$ above $d_c$, (iii) the multiplicative log corrections to the FSS of $\xi$ in a multitude of models at $d_c$ and their absence in models away from $d_c$, we felt it constitutes progress.

Let us look at
the problem with the elaboration of equation~(\ref{1.16}), but before that, a few comments are in order to clarify the meaning of the different lengths introduced in the original submission. The quantity called {\em correlation length}, $\xi(t)\sim|t|^{-\nu}$, is, as usual, the characteristic length appearing in the correlation functions, e.g., that which measures the typical exponential decay of the correlations when approaching criticality. What we denote as $\xi_{Q_L}\sim L$ and call {\em characteristic length} is just the physical length associated with the lattice (Q-space) and  $\xi_{P_L}\sim L^\smallkoppa$, called {\em emergent}, is the finite-size critical correlation length (in P-space). The {\em thermodynamic length} was introduced in reference~\cite{3} to account for the fact that $\xi_{Q_L}$ does not govern FSS in P-space.
In terms of our notations,  it is
$\ell(t)\sim|t|^{-\nu/\smallkoppa}$ and its FSS counterpart
was called {\em coherence length}
 in reference~\cite{19}.

In terms of the finite-size correlation length, the volume of the system in emergent space is $P = (\xi_L/x_0)^{d_c}$ while that of the physical space is $Q = (L/z_0)^{d}$. Here, $x_0$ and $z_0$ are lattice units.

We will show that the hypothesis
\be 
P \sim Q\quad \hbox{or} \quad \xi_L^{d_c} \sim L^d
\label{2.3}
\ee 
 is valid to first approximation (i.e., to the leading order) for
$d_c= 4$, where the symbol ``$\sim$'' indicates asymptotic proportionality (similar scaling behaviour at least to leading order). Equation~(\ref{2.3}) does not constitute a full description of correlation-volume scaling because it does not capture the logarithmic corrections in equation~(\ref{1.14}) which are present when $d = 4$.

Moreover, the map associated with equation~(\ref{2.3}) is not bijective. In particular, if $\koppa > 1$, a given set of coordinates in some reference frame of the P-space is not sufficient to reconstruct a corresponding event in the Q-space unambiguously. The loss of information in going from Q-space to P-space is associated with the fact that long-range correlations are introduced in Q-space automatically. The reason is that even short-range correlations between contiguous sites in P-space translate to interactions between several non-adjacent sites in Q-space. These are consistent with the notion that $\xi_L$ emerges from dangerous irrelevant variables (DIVs) and Renormalization Group (RG) (see reference~\cite{BEHK}) but not the other way --- the RG is a semi-group, not a group. These arguments mean the information-cardinality of Q-space is greater than that of P-space --- a Q-microstate contains more information than a P-microstate.

Therefore, the RG DIV transformation from Q-space to P-space is logically irreversible in the sense that the former cannot be uniquely determined from the latter. According to the Landauer Principle, information is physical and the deletion of information is a dissipative process accompanied by an entropy increase, and conversion of information to energy is possible~\cite{Landaeur,Toyabe,Berut}.
The information content is measured through the Shannon
entropy, 
\be H =-\sum_{i=1}^W
p_i \ln p_i, \ee
where $p_i$ is the probability of the particular message $i$ from the ``message space'' which contains $W$ messages~\cite{Shannon}. If the microstates $i$ have equal probability, the information entropy becomes the Hartley entropy~\cite{Hartley}
\be H = \ln W.\ee
In the case of thermodynamic probabilities, the connection between information entropy and thermodynamic Gibbs entropy comes by identifying the (reduced) Gibbs entropy with the amount of Shannon information needed to uniquely determine the microscopic state of the system from its macroscopic description.

Here, we write the total energy content of Q- and P-space as
\be
E_Q=Q\epsilon_Q,\quad E_P=P\epsilon_P,
\ee
in terms of the energy densities in each space, and these energies are expected to differ due to the difference in their information content. In the discrete case,
$
W={P!}/{(P-Q)!}
$ 
distinct pieces of information (messages) are required to
  reconstruct Q-space from P-space. The corresponding Hartley entropy is
\be
H_{Q\to P}=\ln \frac{P!}{(P-Q)!}.
\ee
The Landauer energy gain associated with this loss of information in going from Q-space to P-space is $k_{\rm B} T H_{Q\to P}$. Conservation of energy/information then demands that
\be
E_P -E_Q =k_{\rm B}TH_{Q\to P}.
\ee
Now, defining $\lambda=\epsilon_P/\epsilon_Q$ and $\kappa = k_{\rm B}T/\epsilon_Q$, we have, using Stirling's formula,
\be \lambda P-Q=\kappa(P\ln P -(P-Q)\ln(P-Q)-Q), \ee
and introducing $r=Q/P<1$ this becomes
\be
\lambda P-Q=\kappa Q(\ln P+(1-r^{-1})\ln(1-r)-1).
\ee
The leading contribution reads as  $\lambda P-Q \simeq \kappa Q\ln P$ [then, $r\simeq 1/(\kappa\ln P)$ and $(1-r^{-1})\ln(1-r)\simeq 1$], which can be iterated to produce
\bea
\lambda P-Q &\simeq& \kappa Q\ln P\nnb\\
&\simeq& \kappa Q(\ln Q+\ln\ln Q+\dots)\nnb\\
&\simeq& \kappa Q\ln Q\Bigl(1+{\cal O}\Bigl(\frac{\ln\ln Q}{\ln Q}\Bigr)\Bigr).\label{2.25}
\eea
Note that the dominant $Q$ term is not that on the left which we started with. Instead, it is the newly arising term coming from the information entropy -- the term on the right. 

At $d=d_c=4$, inserting $P\sim\xi_L^4$  and $Q \sim L^4$ into equation~(\ref{2.25})
recovers equation~(\ref{1.14}) with \be\hat\koppa = 1/d_c=1/4.\ee
To deal with the $d > 4$ case, we have to take the character of the interactions on the two spaces into account. We started with a model in Q-space with only nearest-neighbor interactions. Long-range correlations emerge in Q-space ($\eta_Q < 0$ and $\koppa > 1$) but not in P-space ($\eta_P = 0$). Therefore, the long-ranged correlations are a property of the Q-lattice, not a property of the spin interactions, which remain short-range. To take this into account, we promote the measure of information content to Tsallis’ $q$-entropy~\cite{GellMann}. A minimalist way to do this is to promote the entropic logarithm in equation~(\ref{2.25}) to a $\tilde q$-logarithm:
\be
\lambda P-Q\simeq \kappa Q\ln_{\tilde q} Q,\label{2.26}
\ee
where 
\be
\ln_{\tilde q} Q=\frac{Q^{\tilde q-1}-1}{\tilde q-1}
\ee
which delivers the ordinary logarithm when $\tilde q\to 1$.
Equation~(\ref{2.26})  is then  equation~(\ref{1.16}) in the form
\be
\lambda \xi_L^{d_c}\simeq \kappa L^d\ln_{\tilde q} (L^d)+L^d.
\ee
The value of $\tilde q$ is chosen to recover equation~(\ref{1.2}) to the leading order, as well as the main corrections to scaling calculated by Luijten and Bl\"ote~\cite{BL}. In the case of the susceptibility they got
\be
\chi_L\simeq L^{d/2}(a_0+a_1\hat t L^{d/2}+b_1 L^{4-d}+\dots),
\ee
where $\hat t\sim  L^{4-d-d/2}$, i.e. $\chi_L\sim L^{d/2}(1+cL^{4-d}+\dots)$. In the case of the correlation length, this would translate into 
\be
\xi_L\sim L^{d/4}(1+cL^{4-d}+\dots).
\ee
This is exactly what we get using $\tilde q=1/\koppa$.

\subsection*{The problem with FBCs}

Referee B did not feel ``that a coherent and viable FSS theory has been developed for
free boundary conditions''.
Referee C suggested checking the rounding exponent alongside the other exponents. We had done this, and we confirmed the Referee’s expectation that both the rounding and shifting exponents are $d/2$ for PBCs and for FBCs we confirmed the Referee’s expectation that the shift exponent is 2 and the rounding exponent is $d/2$.
The Referee’s motivation here was that, if the shift is bigger than the rounding, $T_c$ will be too far from $T_L$ to ``feel'' the peak --- which will be outside the FSS zone. This would explain why, even if $\chi_L(T_L)$ (at the pseudocritical point) scales as $L^{d/2}$ for FBCs, $\chi_L(T_c)$ (at the critical point) may scale differently, like $L^2$ at $T_c$, and may rescue Gaussian FSS at $T_c$ for FBC. 

We had checked this explicitly too and we found that in 5D $\chi_L(T_c)$ (at the critical point with FBCs) scales as $L^{1.71(2)}$ using all sites of the Q-lattices or $L^{1.92(2)}$ using only sites at the core of the lattices. This was close to, but not quite the Gaussian behaviour expected by the Referee. 
Therefore, we had not been able to confirm the Referee’s expectation that it is Gaussian at criticality. 
 Specifically, we were able to claim, in our revised version that
(i) Q-space systems with PBCs and $d > d_c$ are not Gaussian either at criticality or at pseudocriticality. Instead $\koppa = d/d_c$ governs modified FSS. This was in agreement with Luijten’s and Bl\"ote’s numerics in the PBC case at criticality~\cite{Lthesis}.
(ii) Q-space $d > d_c$ systems with FBCs are not Gaussian at the pseudocritical point. Instead, $\koppa = d/d_c$ governs modified FSS there. Thus, pseudocriticality with FBCs is essentially the same as pseudocriticality with PBCs and obeys modified FSS.
(iii) Q-space $d > d_c$ systems with FBCs may or may not be Gaussian at the critical point, but with currently available lattices our numerics were not supportive of Gaussian behaviour there. The question went on attracting the attention of the community later~\cite{LM,Emilio}, and even recently, Ralph was still involved in trying to solve this difficult question~\cite{Julian}.
(iv) All $d > d_c$ systems in P-space are Gaussian, as per Landau theory.

\section*{Since 2012}
Since 2012, a few papers that I consider as important in the field have been published, clarifying, or completing some of the perspectives presented in the present paper.

As far as I know, Wittmann and Young~\cite{Wittmann} were the first to study the FSS of Fourier modes in high dimensional Ising models. They have shown that the modified FSS that  allows for violation of hyperscaling due to a dangerous irrelevant variable applies only to ${\bf k} = 0$ fluctuations, and that ``standard'' FSS applies to ${\bf k}\not= 0$ fluctuations. Nevertheless, the denomination of ``standard'' was referring to Landau scaling while an elucidation in references~\cite{Emilio,Emilio2} has shown that this should be understood as Gaussian Fixed Point scaling. 

The universality class of the percolation problem above its upper critical dimension $d_c=6$ was studied from the perspective of the role of the dangerous irrelevant variable in systems with PBCs and FBCs in reference~\cite{Perco}.

After the spectacular work of Luijten~\cite{Lthesis} in 1997, 
the case of the logarithmic corrections of the Ising Model exactly at $d_c=4$  was revisited very convincingly recently by Lv et al.~\cite{Deng}.

The work of Langheld et al.~\cite{Langheld} has extended Q-FSS to quantum systems in a remarkable manner. The study was performed in the case of a finite $d$-dimensional Transverse Field Ising Model (TFIM), a quantum system of Pauli spin operators interacting in nearest neighbours, $\sim\sigma_i^z\sigma_j^z$ (the long-range interacting case was also considered), with an additional transverse interaction $\sim h\sigma_i^x$. The quantum time evolution plays the role of an additional space dimension for the classical analogue $L^d\times\infty$ of dimension  $D=d+1$. For the $d$-dimensional quantum system, the upper critical dimension is thus $d_c=3=D_c-1$ and the exponent $\koppa$ takes the value $\koppa= d/3={(D-1)}/{(D_c-1)}$ which differs from a $D$-dimensional classical analogue for which one would have 
$\koppa_{\rm cl}=D/4$. The hyperscaling  relation also needs to be rewritten. $d$-dimensional classical systems below their upper critical dimension have $2-\alpha=\nu d$. $d-$dimensional quantum systems have $2-\alpha=\nu(d+z)$ in terms of the anisotropy exponent (that distinguishes the time direction from the spatial ones), and above their upper critical dimension, quantum systems have $2-\alpha=\nu(\frac d\smallkoppa+z)$.

Random field Ising models above their upper critical dimension ($d_c=6$) in $d=7$ were studied very recently by Fytas et al. in reference~\cite{Fytas} where the anomalous FSS of the correlation length $\xi_L\sim L^{\frac 76}$ was confirmed for the first time in a disordered system.

\section*{Some personal thoughts}
A scientist leaves a mark through his scientific production, and his articles, but also through how he has left his mark on the people he has encountered, his students, and his colleagues. Ralph is certainly one of those people whose memory will stay with us for a long time. I think he would have been happy to have this article published and I don't think I'm betraying his wishes by submitting it to Condensed Matter Physics.

\section*{Acknowledgement}
I thank Jean-Charles Walter for having accepted the publication of this previously unpublished paper, Thierry Platini for the wonderful photo, and my friends Olesya, Reinhard and Yurko,  the editors of this issue for their support. I thank Ernesto Medina for the comforting discussions we had during this period full of contradictory emotions and Tim Ellis for his nice words.
Of course, I also want to express here my deepest support for Claire and Ro\'\i s\'\i n.  Ro\'\i s\'\i n, you can be proud of what your father has accomplished.

\ukrainianpart

\title{Про одну неопубліковану роботу з Ральфом Кенною}
\author{\framebox{Р. Кенна}\refaddr{label1,label3}, Б. Берш\refaddr{label2,label3}}
\addresses{
	\addr{label1} Центр дослідження рідин і складних систем, Університет Ковентрі, Англія
	\addr{label2} 
	Лабораторія теоретичної фізики та хімії, CNRS -- Університет Лотарингії, UMR 7019,
	Нансі, Франція 
	\addr{label3} Колаборація $\mathbb{L}^4$, Ляйпциг-Лотарингія-Львів-Ковентрі, Європа
}

\makeukrtitle

\begin{abstract}
Це частина неопублікованої роботи у співпраці з Ральфом Кенною. Ймовірно, вона була ще недостатньо зрілою на момент подання понад десять років тому. Відтак, її відхилили редактори, але деякі ідеї пізніше були частково опубліковані у наступних роботах. Я вважаю, що ця ``чернетка'' дуже багато розповідає про ентузіазм та сміливість Ральфа і тому заслуговує на публікацію зараз, можливо, як частина його спадщини.

\keywords 
універсальність, скінченновимірний скейлінг, верхня критична розмірність
\end{abstract}
\lastpage
\end{document}